\documentclass[twocolumn,prl,superscriptaddress]{revtex4}
\usepackage[latin9]{inputenc}
\usepackage{color}
\usepackage{amsmath}
\usepackage{amsthm}
\usepackage{amssymb}
\usepackage{graphicx}
\usepackage{float}

\usepackage{algorithm}  
\usepackage{algorithmicx}  
\usepackage{algpseudocode}

\usepackage[unicode=true,
 bookmarks=true,bookmarksnumbered=false,bookmarksopen=false,
 breaklinks=false,pdfborder={0 0 1},backref=false,colorlinks=true]
 {hyperref}
\hypersetup{
 linkcolor=magenta, urlcolor=blue, citecolor=blue, pdfstartview={FitH}, hyperfootnotes=false, unicode=true}

\makeatletter
\@ifundefined{textcolor}{}
{%
 \definecolor{BLACK}{gray}{0}
 \definecolor{WHITE}{gray}{1}
 \definecolor{RED}{rgb}{1,0,0}
 \definecolor{GREEN}{rgb}{0,1,0}
 \definecolor{BLUE}{rgb}{0,0,1}
 \definecolor{CYAN}{cmyk}{1,0,0,0}
 \definecolor{MAGENTA}{cmyk}{0,1,0,0}
 \definecolor{YELLOW}{cmyk}{0,0,1,0}
}

\usepackage{amsfonts,amsthm}\usepackage{tabularx}\usepackage{dcolumn}\usepackage{bm}\usepackage{graphicx}\usepackage{epstopdf}
\usepackage{times}

\hypersetup{urlcolor=blue}

\def\RR{{\mathbb R}}

\def\HH{{\mathcal H}}
\def\OO{{\mathcal O}}
\def\EE{{\mathcal E}}

\def\UU{{\mathcal U}}

\makeatother

\begin{document}

\title{Weighted Quantum Channel Compiling through Proximal Policy Optimization}
\author{Weiyuan Gong}
\thanks{These authors contributed equally to this work.}
\affiliation{Center for Quantum Information, IIIS, Tsinghua University, Beijing 100084, China}
\author{Si Jiang}
\thanks{These authors contributed equally to this work.}
\affiliation{Center for Quantum Information, IIIS, Tsinghua University, Beijing 100084, China}
\author{Dong-Ling Deng}
\email{dldeng@tsinghua.edu.cn}
\affiliation{Center for Quantum Information, IIIS, Tsinghua University, Beijing 100084, China}
\affiliation{Shanghai Qi Zhi Institute, 41st Floor, AI Tower, No. 701 Yunjin Road, Xuhui District, Shanghai 200232, China}
%

\begin{abstract}
We propose a general and systematic strategy to compile arbitrary quantum channels without using ancillary qubits, based on proximal policy optimization---a powerful deep reinforcement learning algorithm. We rigorously prove that, in sharp contrast to the case of compiling unitary gates, it is impossible to compile an arbitrary channel to arbitrary precision with any given finite elementary channel set, regardless of the length of the decomposition sequence. However, for a fixed accuracy $\epsilon$ one can construct a universal set with constant number of $\epsilon$-dependent elementary channels, such that an arbitrary quantum channel can be decomposed into a sequence of these elementary channels followed by a unitary gate, with the sequence length bounded by $O(\frac{1}{\epsilon}\log\frac{1}{\epsilon})$. Through a concrete example concerning topological compiling of Majorana fermions, we show that our proposed algorithm can conveniently and effectively reduce the use of expensive elementary gates through adding the weighted cost into the reward function of the proximal policy optimization.
\end{abstract}
\maketitle

\begin{figure}
    \centering
    \includegraphics[width=0.481\textwidth]{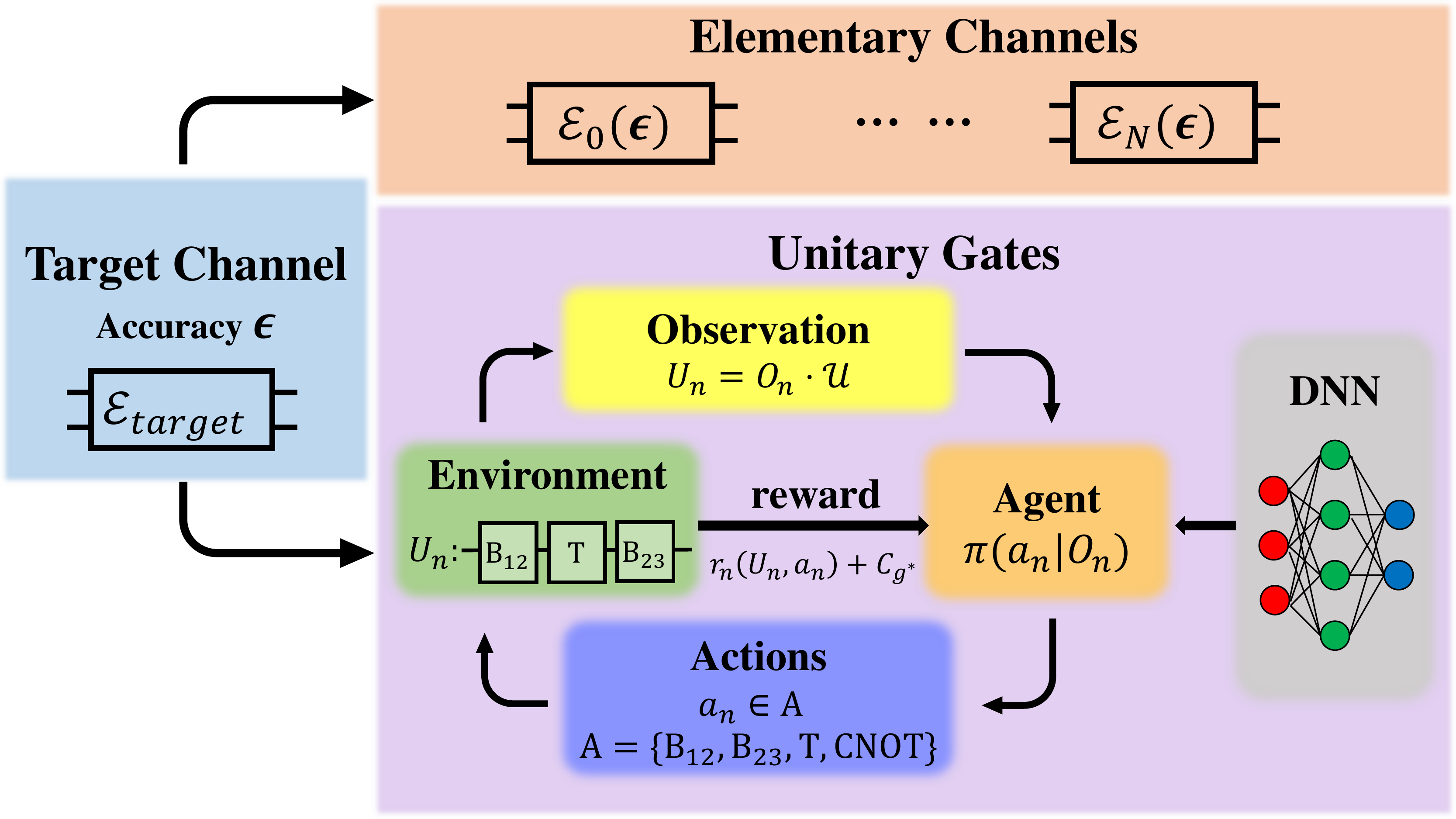}
    \caption{A schematic illustration of the quantum channel compiler based on deep reinforcement learning (DRL). The compiler constructs a constant size elementary channel set $\EE_0(\epsilon),...,\EE_N(\epsilon)$ using $\epsilon$ as parameters for each channel. It first decomposes the target channel into a sequence of elementary channels followed by a unitary gate. Then a DRL environment is set up to produce approximation sequence $U_n$ for the target gate $\UU$ required in the previous step using elementary gates. At each step $n$, the agent gets an observation $O_n$ and feeds $O_n$ as the input vector to the deep neural network (DNN). DNN outputs a probability distribution $\pi(a_n|O_n)$, according to which the agent chooses the next gate to apply in the decomposition sequence. The environment returns a reward $r_n$ to the agent afterward. See the Supplementary Materials for details \cite{supplement}.}
    \label{fig:ProcessIntro}
\end{figure}

Quantum compilers, which decompose quantum operations into hardware compatible elementary operations, play an important role in quantum computation \cite{Nielsen2010Quantum} and digital simulation \cite{Georgescu2014Quantum}.
This technique is especially crucial for the applications of noisy intermediate-scale quantum devices \cite{Preskill2018Quantum}, where the performance of deep quantum circuits might be limited by noises and quantum decoherences. A number of notable approaches have been proposed to compile unitary gates and the dynamics of isolated quantum systems \cite{Dawson2005Solovay,Kitaev2002Classical,Jones2012Faster,Fowler2011Constructing,Bocharov2012Resource,Bocharov2013Efficient,Pham2013Optimization,Zhiyenbayev2018Quantum,Kliuchnikov2013Asymptotically,Selinger2013Quantum,Gosset2014Algorithm,Ross2014Optimal,Heyfron2018Efficient}. However, in reality quantum systems cannot be perfectly isolated and would inevitably interact with the external environment, making the more general quantum channel compiling indispensable for a wide range of applications \cite{Breuer2002Theory,Lidar2019Lecture}.
Yet, quantum channel compilation has been barely explored \cite{Braun2014Universal}, with major previous attention paid to exploiting the Stinespring dilation theorem \cite{Stinespring1955Positive} and compiling arbitrary quantum channels through elementary gates acting on an expanded Hilbert space with ancillary qubits playing an prerequisite role \cite{Wang2013Solovay,Iten2017Quantum,Wei2018Efficient,Iten2016Quantum,Shen2017Quantum,Sweke2014Simulation,Passos2020Spin,Wang2015Quantum,Lu2017Experimental,Xin2017Quantum}. Hitherto, a general and systematic strategy to compile arbitrary quantum channels without using ancillary qubits has not been established. Here, we prove two generic theorems regarding channel compiling and introduce such a strategy based on deep reinforcement learning (see Fig. \ref{fig:ProcessIntro} for an illustration).

Machine learning, or more broadly artificial intelligence, has recently cracked a number of notoriously challenging problems, such as playing the game of Go \cite{Silver2016Mastering,Silver2017Mastering}, predicting protein spatial structures \cite{Senior2020Improved}, and weather forecasting \cite{Ravuri2021Skilful}. Its tools and techniques have been broadly exploited in various quantum physics tasks, including representing quantum many-body states \cite{Carleo2016Solving,Gao2017anEfficient}, quantum state tomography \cite{Torlai2018Neural,Carrasquilla2019Reconstructing}, learning topological phases of matter \cite{Zhang2017Quantum,Carrasquilla2017Machine,vanNieuwenburg2017Learning,Wang2016Discovering,Broecker2017Machine,Chng2017Machine,Zhang2017Machine,Wetzel2017Unsupervised,Hu2017Discovering,Zhang2019Machine,Lian2019Machine}, and nonlocality detection \cite{Deng2017MachineBN}. For quantum compiling on unitary gates, machine learning approaches have also been introduced to provide a near-optimal sequence \cite{Alam2019Quantum,Zhang2020Topological}. In this paper, we first rigorously prove that it is impossible to compile any quantum channel to arbitrary accuracy using unitary gates and a finite set of elementary channels, which is in sharp contrast to the case of compiling a unitary gate. As illustrated in Fig. \ref{fig:ProcessIntro}, we propose a quantum channel compiler which given an accuracy demand $\epsilon$, decomposes any quantum channel into a sequence of finite types of elementary quantum channels followed by a unitary gate. We provide a constructive method to obtain the elementary channel set and show that the size of the set scales as $O(d^2)$ with the dimension $d$ of Hilbert space and is independent of $\epsilon$. We additionally prove that the length of the elementary channel sequence in the decomposition is bounded above by $O(\frac1\epsilon\log(\frac1\epsilon))$. For the unitary gate at the end of the decomposition, we train a deep reinforcement learning (DRL) agent to decompose it into hardware compatible elementary gates. To reduce the resource requirement of the compiler, we exploit the proximal policy optimization (PPO) algorithm \cite{Schulman2017Proximal} to train our agent with weighted cost  in the reward function to reduce the use of experimentally-expensive elementary gates. We further prove a $\Omega(\log(1/\epsilon))$ lower bound for any indispensable expensive  elementary gate count to compile an arbitrary unitary gate within error $\epsilon$. As a benchmark, we apply our algorithm to the topological quantum compiling of Majorana fermions \cite{Kitaev2006Anyons,Nayak2008Non}, whose braidings together with a non-topological $T$ gate form a universal set. 
We numerically show that our algorithm could reduce the use of $T$ gate by a factor of two compared to the traditional Solovay-Kitaev algorithm.

\textit{Notations}.---To begin with, we first introduce some basic notations and concepts \cite{Nielsen2010Quantum}. A quantum state can be represented by a positive semi-definite operator $\rho\in\OO(\HH_S)$ with $\text{Tr}(\rho)=1$, where $\HH_S$ is the Hilbert space and $\OO(\HH_S)$ the set of operators on $\HH_S$. In general, a quantum channel $\EE$ can be characterized by a completely positive, trace-preserving (CPTP) map which maps a quantum state $\rho$ into another state $\EE(\rho)\in\OO(\HH_S)$. Any single-qubit state can be represented as $\rho=\frac12(I+\bm{a}\cdot\bm{\sigma})$, where $\bm{a}$ is a three-dimensional vector within the Bloch sphere and $\bm{\sigma}=(\sigma_x,\sigma_y,\sigma_z)$ are Pauli matrices. Any linear CPTP map for a single-qubit system  could be represented by a four-by-four matrix $\mathcal T$ \cite{Ruskai2002Analysis,Wolf2008Dividing,Wang2013Solovay,Wang2015Algorithmic}:
\begin{equation}\label{eq:channel_matrix_1}
\EE\to \mathcal T=\begin{pmatrix}
1 & 0\\ t & T
\end{pmatrix},\mathcal{T}_{ij}=\frac12\text{Tr}[\sigma_i\EE(\sigma_j)],
\end{equation}
where $i,j\in \{0,1,2,3\}$, $\sigma_0,\sigma_1,\sigma_2,\sigma_3$ represents the Pauli matrix $I,\sigma_x,\sigma_y,\sigma_z$ and $T\in\RR^{3\times3},t\in\RR^3$. Under this representation, a channel is an affine map \cite{King2001Minimal} $\EE:\frac12(1+\bm{a}\cdot\bm{\sigma})\to\frac12(1+\bm{a}'\cdot\bm{\sigma}),\bm{a}'=T\bm{a}+t$. Geometrically, $\EE$ maps the states within the Bloch sphere into states enveloped by an ellipsoid, with $t$ the center shift from the original center and $T$ the distortion matrix for the ellipsoid. When $\det (T)=1$, the CPTP map reduces to a unitary gate. In this sense, unitary gates can be regarded as special channels. Throughout this paper, we differentiate unitary gates from channels for clarity. 

\textit{A general theorem for channel compilation.}---To formulate the problem, we consider a set $S$ with metric $d(\cdot)$. A set $\Gamma\subset S$ is called a $\delta$-net if for any $x\in S$, there exists $y\in \Gamma$ such that $d(x,y)\leq\delta$ \cite{Kitaev2002Classical}.  The subset $\Gamma\subset S$ is called a dense subset under metric $d(\cdot)$ if it is a $\delta$-net of $S$ for arbitrary $\delta$.

Suppose we have a set of elementary channels and want to approximate the target channel with a sequence of unitary gates and elementary channels chosen from the set. For technical convenience and simplicity, we consider Schatten one-norm \cite{Kliesch2011Dissipative} $||\EE_{\text{target}}-\EE_{\text{approx}}||_{1\to1}=\max_{\rho\in\OO(\HH_S)}||\EE_{\text{target}}(\rho)-\EE_{\text{approx}}(\rho)||_1$ as the distance measure. Now, we are ready to present our general theorem.




\textbf{Theorem 1. }Consider compiling single-qubit channels using unitary gates and elementary channels with Schatten one-norm as the distance measure. Then:

(1) Given a finite set of  elementary channels together with an arbitrary unitary gate, it is impossible to compile an arbitrary single-qubit channel to arbitrary accuracy.

(2) Given an accuracy bound $\epsilon$, one can construct a finite set of elementary quantum channels using $\epsilon$ as a parameter such that any single-qubit channel can be compiled by the elementary channels from this set and a unitary gate within error $\epsilon$. The length of sequence is bounded above by $O(\frac1\epsilon\log(\frac1\epsilon))$.

\textit{proof.} We provide the main idea here. The full proof is technically involved and thus left  to the Supplementary Materials \cite{supplement}. Suppose that we are provided with a finite set of  elementary channels $\mathcal{C}=\{\EE_1,...,\EE_n\}$ with corresponding distortion matrices $\{T_1,..,T_n\}$ and center shifts $\{t_1,...,t_n\}$. Without loss of generality, we assume that $\det{(T_n)}\leq...\leq\det{(T_1)}<1$. 
Noting that the composition of channels could not increase the determinant of the distortion matrix, thus a target channel with  $\det{(T_1)}<\det{(T)}$ cannot be compiled by the channels chosen from $\mathcal{C}$ to arbitrary accuracy, independent of how long the decomposition sequence is. 
For part (2), we decompose the compiling process into several steps and provide a constructive proof. For a target channel $\EE$ with distortion matrix $T$ with eigenvalues $\lambda_1,\lambda_2,\lambda_3$ and center shift $t$, we first implement intermediate channel $\EE_1$ with parameters $T_1=\text{diag}(|\lambda_1|,|\lambda_2|,|\lambda_3|), t_1=t$ using elementary channels parametrized by $\epsilon$. We then use a unitary gate to realize the negative, complex eigenvalues and basis transformations. 

The above theorem could be extended to multi-qubit channels. For a $d$-dimensional quantum state $\rho\in\OO(\HH_S)$, there exists a canonical and orthonormal basis $\{O_\alpha, \alpha=1,...,d^2-1\}$ \cite{Wang2015Algorithmic,Bruning2012Parametrizations}. A density operator $\rho$ under such basis could be written as $\rho=\frac 1d(I+\sum_{\alpha=1}^{d^2-1}p_\alpha Q_\alpha),Q_\alpha=\sqrt{d(d-1)}O_\alpha$. The parameters in $\{p_\alpha\}$ form the polarization vector $\bm{p}=(p_1,...,p_{d^2-1})$ of a $(d^2-1)$-dimensional ball with $||p||_2=1$ representing pure states and $||p||_2<1$ representing mixed states. As a quantum state $\rho$ can be represented by a vector within a ball, a quantum channel $\EE:\OO(\HH_S)\to\OO(\HH_S)$ can be written as an affine map represented by distortion matrix $T\in\RR^{(d^2-1)\times (d^2-1)}$ and center shift $t\in\RR^{d^2-1}$ similar to Eq.~\eqref{eq:channel_matrix_1}. With this representation, we can extend the Theorem 1 to the multi-qubit case \cite{supplement}. Yet, it is worthwhile to mention that in this case, the size of set of constructed quantum channels should scale as $O(d^2)$ in part (2) of the theorem \cite{supplement}.



The above results imply that a finite number of elementary channels could not approximate an arbitrary target channel to arbitrary accuracy, regardless of the specific structure of each elementary channel and the length of the compiling sequence. This is in sharp contrast with the case of unitary gate compiling, where we can use a small number of elementary gates to compile an arbitrary unitary gate within any accuracy demand. We remark that any quantum channel can be implemented by a sequence of elementary unitary gates acting on a dilated Hilbert space and this seems to contradict with the claim of part (1) in the Theorem 1. However, this spurious contradiction dissolves after noting the fact that tracing out the ancillary qubits at different sequence locations would effectively result in different channels even for the same elementary unitary gates. In other words, although a small number of different unitary gates suffice to implement any quantum channel with ancillary qubits, when restricted to the targeted system \textit{no} finite set of elementary channels is universal.


In the proof for part (2) of the Theorem 1, we have provided two explicit constructions to decompose an arbitrary quantum channel into a sequence of elementary channels followed by a unitary gate \cite{supplement}. The first construction has an elementary channel set of $O(d^2)$ size with a sequence length $O(d^2/\epsilon\log(1/\epsilon))$, and the other uses a much larger elementary channel set [of size $O(2^{d^2})$] but much shorter decomposition sequence [of length $O(1/\epsilon\log(1/\epsilon))$]. In other words, we can decompose any target quantum channel into a fixed sequence of  elementary channels followed by a $n$-qubit unitary gate. The channel compilation task has thus been reduced to unitary compiling with elementary unitary gates.


\textit{Weighted unitary gate compilation and a DRL algorithm}.---We now consider quantum compilation for unitary gates $U\in SU(d)$ with the elementary gate set $S_E=\{g_1,...,g_n|g_i\in SU(d)\}$. A gate set is universal if it can compile arbitrary unitary gates to any given accuracy demand under the distance measure. In other words, a gate set is universal if and only if it generates a dense subgroup in $SU(d)$ \cite{Kitaev2002Classical}. 
We present the following theorem concerning the lower bound of any indispensable gate in compiling an arbitrary unitary.

\textbf{Theorem 2. }For a non-dense subgroup $G\subset SU(d)$ generated by $S_E$, suppose we can find $g^*\in SU(d)$ such that $G'$ generated by $\{g_1,...,g_n,g^*\}$ is dense in $SU(d)$. When employing $G'$ as elementary gate set for quantum compilation task on $SU(d)$, the number of gate $g^*$ to compile an arbitrary gate within distance $\epsilon$ is bounded below by:
\begin{equation}\label{eq:compile_lower}
N^*=\Omega(\frac{d^2-1}{\log(|G|)}\log(\frac1\epsilon)).
\end{equation}

The proof of Theorem 2 relies on the volume method, which considers covering the whole $SU(d)$ space with $\epsilon$-balls centered at each possible gate sequence. For brevity, we leave the technical details to the Supplementary Materials \cite{supplement}. This theorem gives a lower bound for the count of any indispensable gate $g^*$ in compiling an arbitrary unitary, which scales linearly in $\log (\frac{1}{\epsilon})$ but quadratic in the Hilbert space dimension $d$ that is exponentially large as the system size increases. In practical applications, $g^*$ may represent some experimentally expensive or flawed gate and thus reducing its count in compiling could be crucial. For the case of quantum compiling with Clifford$+T$ gate set, a number of striking algorithms \cite{Kliuchnikov2013Asymptotically,Selinger2013Quantum,Gosset2014Algorithm,Ross2014Optimal,Heyfron2018Efficient}, which either exploit its specific structures or utilize ancillary qubits, have been proposed to reduce the $T$ count. Here, we introduce a more general approach (in the sense that it does not rely on the special properties of the elementary gate set and thus bears universal applicability) without using ancillary qubits. We exploit a reinforcement learning technique, the proximal policy optimization  \cite{Schulman2017Proximal} in particular,  to reduce the count of experimentally expensive gates.




Unlike commonly used Q-learning algorithms \cite{Watkins1992Q,Zhang2020Topological} such as deep-Q networks \cite{Sutton2018Reinforcement}, PPO directly represents a policy explicitly as $\pi_{\theta}(a|s)$ by a neural network, which receives the current state $s$ as an input and outputs the probability $\pi_{\theta}(a|s)$ the agent may choose for each action $a$. 
The updating rules used by PPO explore the biggest possible improvement step without causing a performance collapse \cite{Schulman2017Proximal}. This property makes PPO particularly suitable for quantum compilation tasks since applying an inappropriate gate in the sequence will dramatically destruct the approximation gate. Moreover, we modify the reward function used in PPO, which efficiently reduces the count of a specific gate that is experimentally costly:
\begin{equation}\label{eq:reward}
r_n=r(U_n, U_t) - C_{g^*},
\end{equation}
where $r_n$ is the reward that the agent receives for the $n$-th step, $r(U_n, U_t)$ is the reward determined by comparing the approximation gate $U_n$ and the target gate $U_t$, and $C_{g^*}$ is the additional weighted punishment for the employment of $g^*$ gates. By increasing $C_{g^*}$, the agent would tend to avoid using $g^*$ gates and thus the $g^*$ count would be reduced in the decomposition \cite{supplement}. 
By exploiting the PPO algorithm, our algorithm searches the approximation sequence in a depth-first search scheme. Comparing with the breadth-first search DRL algorithm proposed in Ref. \cite{Zhang2020Topological}, the PPO algorithm runs with significantly less time and memory in both the training and searching stages. This advantage makes our algorithm feasible for compiling tasks for larger systems with more complicated actions and rewards.
\begin{figure}
    \centering
    \includegraphics[width=0.48\textwidth]{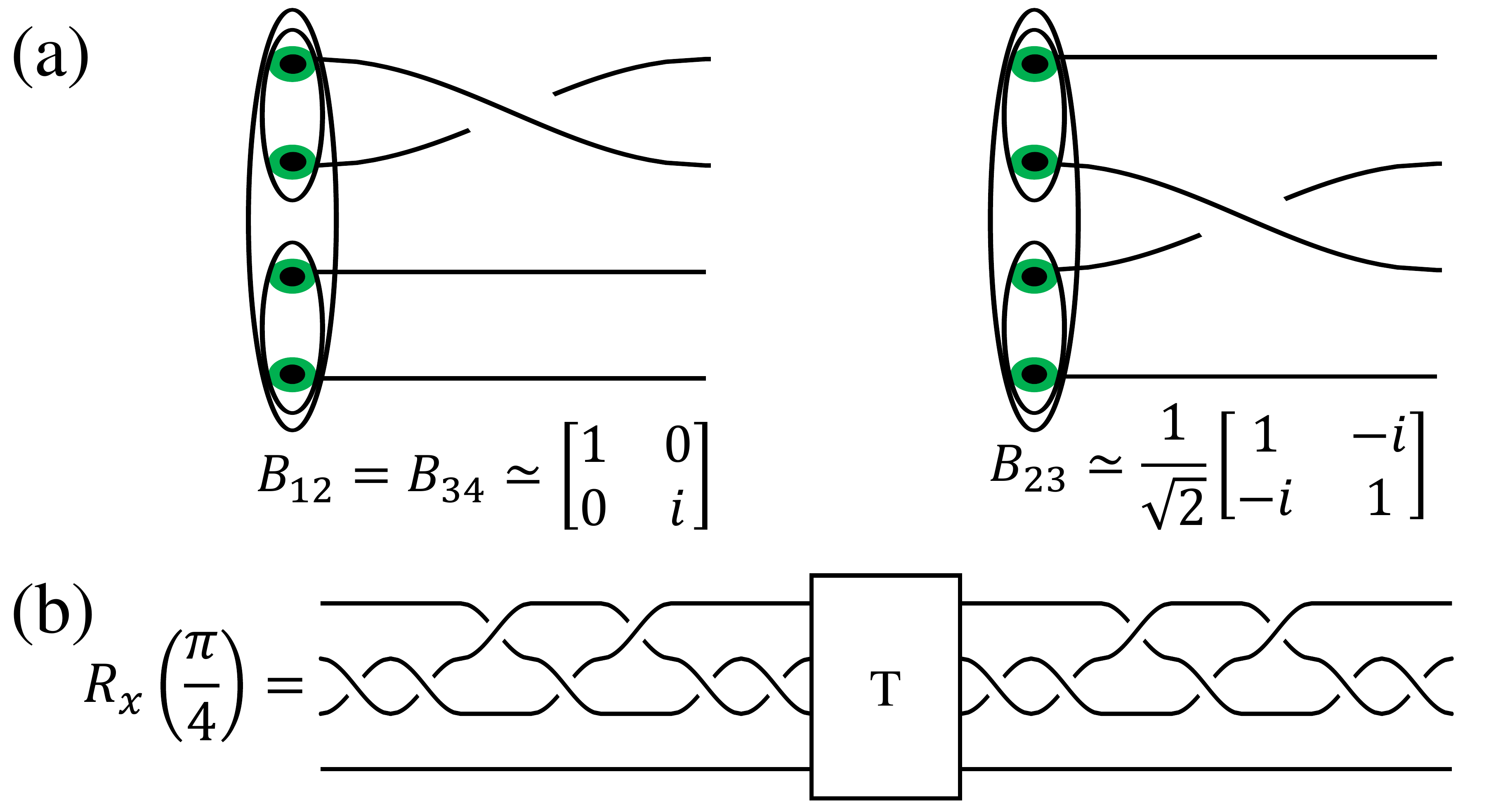}
    \caption{(a) The two elementary gates that can be implemented through one braiding of Majorana fermions. Logical qubits are encoded into four Majorana fermions (enclosed in ovals). (b) An exact decomposition sequence of using Majorana braidings and a $T$ gate for compiling the $R_x(\frac{\pi}{4})$ gate. }
    \label{fig:BraidIntro}
\end{figure}

\textit{Topological compiling of Majorana fermions.} To benchmark the performance of our PPO algorithm in practice, we consider topological compiling with the four-quasiparticle encoding scheme for Majorana fermions \cite{Nayak2008Non}. Two elementary gates corresponding to the braidings of four Majorana fermions are shown in Fig. \ref{fig:BraidIntro}(a) and unitary gates can be approximated through braiding sequences and $T$ gates. A simple example for decomposing an $x$-rotation is  shown in Fig. \ref{fig:BraidIntro}(b). It is well known that braidings of Majorana fermions only lead to an elementary gate set $S_E=\{$CNOT, H, S$\}$, which is not sufficient for universal quantum computation \cite{Bravyi2005Universal,Nayak2008Non}. To achieve universality, a non-topological $T$ gate with a relatively high experimental cost is necessary. Therefore, reducing $T$ count is of practical importance \cite{Gheorghiu2021TCount,Selinger2015Efficient,Matsumoto2008Representation}. Here, we apply the PPO algorithm to attain this goal. To this end, we note that the topological gate set generates a Clifford group $\text{Cl}_n\subset SU(d)$ with finite size $|\text{Cl}_n|=O(2^{2n^2+3n})$ \cite{Calderbank1998Quantum,Nebe2001Invariants,Planat2008Group}. According to the Theorem 2,   the scaling of $T$ count for compiling an arbitrary gate in the worst case is $\Omega(\frac{d^2-1}{n^2}\log(\frac1\epsilon))$.



We mention that topological quantum compiling has been broadly explored with various algorithms proposed \cite{deng2010fault,Bonesteel2005Braid,xu2008Constructing,Hormozi2007Topological,Burrello2010Topological,Kliuchnikov2014Asymptotically,Carnahan2016Systematically}. Most of the algorithms run in $O(\text{poly}\log(1/\epsilon))$ and output sequences of braidings with length $O(\text{poly}\log(1/\epsilon))$ to obtain an approximation within $\epsilon$ distance from given target evolution. Here, we exploit the average gate fidelity $F(U,V)=\int d|\psi\rangle|\langle\psi|UV^\dagger|\psi\rangle|^2$ provided in the open-source QuTiP package \cite{Johansson2012Qutip} to measure the distance $d(U,V)=1-F(U,V)$ between an approximated unitary gate $V$ and a target unitary gate $U$. 
To exploit or DRL algorithm as a single qubit quantum compiler, the action space is a set $A=\{B_{12},B_{12}^{-1},B_{23},B_{23}^{-1},T, T^{-1}\}$ containing six elementary gates. To train the agent, we employ a deep neural network (DNN) with five layers of fully connected neurons and train it with a PPO algorithm encapsulated in \textsc{OpenAI} gym and baseline package \cite{Brockman2016Openai}. To construct the training data, we generate a sequence of length $L$ consisting of the elementary gates to be the target gate. To test our agent, we construct a test dataset of $1500$ data samples, each as a random sequence of gates from $A$ of length between $10$ and $80$. We compare the performance of our algorithm and the traditional Solovay-Kitaev algorithm on such a dataset \cite{supplement}. 
\begin{figure}
    \centering
    \includegraphics[width=0.47\textwidth]{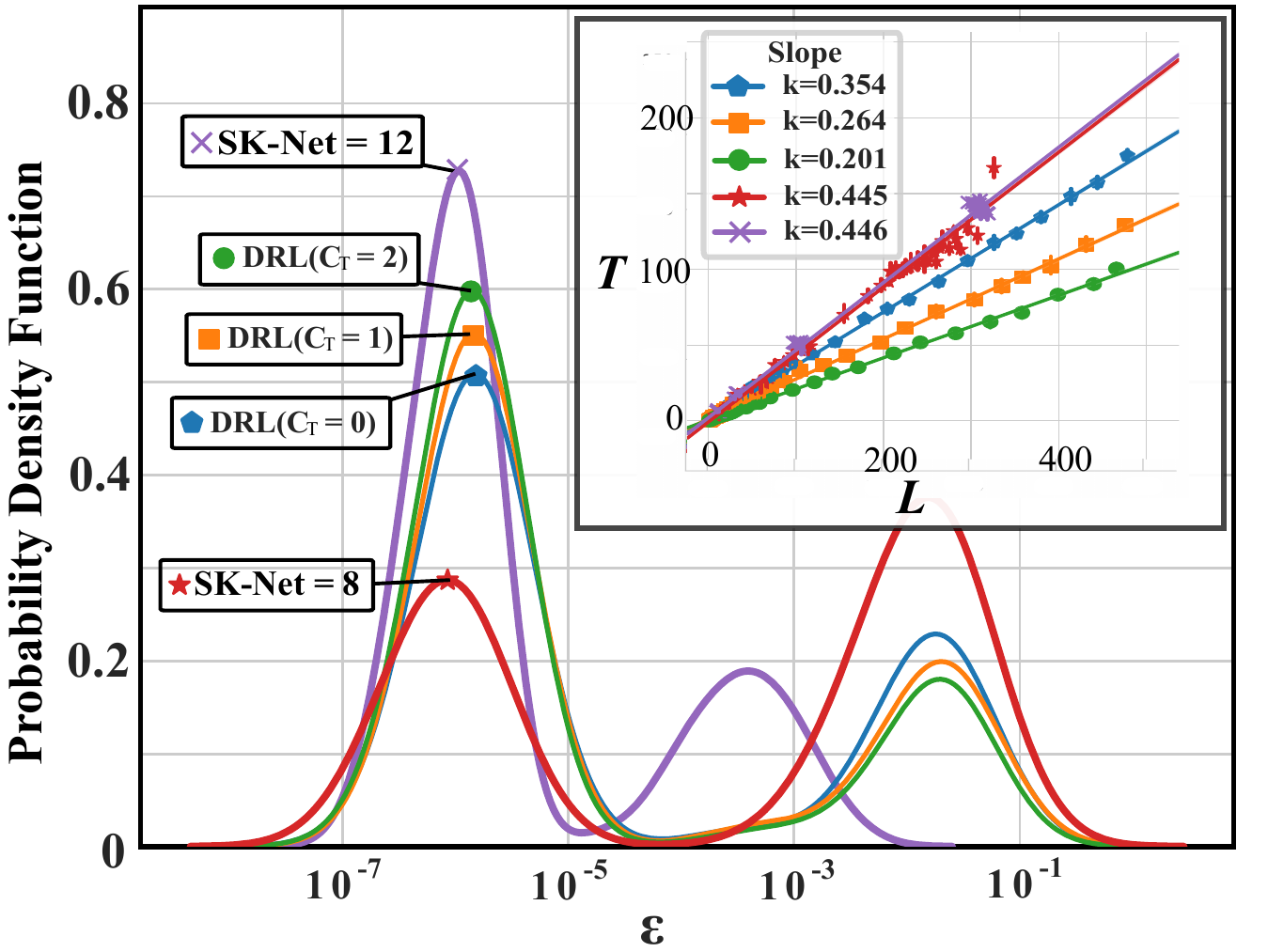}
    \caption{Comparison of the distribution of the distance $\epsilon=d(U_t,U_{\text{approx}})$ between the target gate $U_t$ and the approximation gate $U_{\text{approx}}$ for our algorithm with different $T$ cost $C_T=0,1,2$ and the Solovay-Kitaev (SK) algorithm with net of depth $8$ and $12$. The inset shows the proportion of $T$ gates in the compilation gate sequence for our algorithm and the SK algorithm. The slope $k$ is calculated by fitting the scatter points over all test samples.}
    \label{fig:TRateLengthComp}
\end{figure}

In Fig. \ref{fig:TRateLengthComp}, we plot the error distribution over the test dataset and the proportion of $T$ gate for our algorithm and Solovay-Kitaev algorithm with the same threshold $\epsilon_t=10^{-3}$. 
We observe that the error distributions for both algorithms show two peaks with one below the threshold $\epsilon_t$ and the other larger than the threshold.
This could be understood as the potential risk of failure for the depth-first search scheme, which is exploited in both algorithms. On average, the Solovay-Kitaev algorithm with net depth $12$ provides a more accurate approximation than our algorithm at the price of longer average length while the $8$-depth algorithm provides an equal average length with sacrificed approximation accuracy. In the inset of Fig. \ref{fig:TRateLengthComp}, we find that by increasing the cost of $T$ gate from $C_T=0$ to $C_T=2$, the average $T$ gate rate over the test set decreases from $0.354$ to $0.201$. This is a significant $60\%$ reduction compared to that for the Solovay-Kitaev algorithm which gives a $T$ gate rate of $0.445$. This shows that adding the cost of $T$ gate as a punishment in DRL could effectively reduce the $T$ count. 



We remark that the depth-first search scheme in our PPO algorithm makes the time complexity scale linearly with the maximal search depth $L_{\text{max}}$. This is distinct from the $A^*$ search (which is breadth-first and hence exhibits an exponential scaling with $L_{\text{max}}$) used in Ref. \cite{Zhang2020Topological}. As a result, the PPO algorithm is significantly less time and memory consuming in both the training and searching stages. This advantage makes our algorithm feasible for compiling tasks for larger systems with more complicated actions and rewards. As a trade-off, the PPO algorithm would not output the near-optimal sequence for a given target unitary and accuracy demand, and may even fail to find a decomposition if the accuracy threshold is too small. We mention that one can increase the successful rate of the compilation by increasing $L_{\text{max}}$ \cite{supplement}. 

\textit{Discussion and conclusion}.---Theorem 1 implies that for channel compiling, there exists \textit{no} finite universal elementary channel set. However, for a given target channel, this theorem does not tell whether it can be decomposed into predetermined elementary channels to a desired accuracy. Finding a general and efficiently computable criterion for determining whether a given channel can be compiled with a fixed elementary channel set or not is of both theoretical and experimental importance, and worth future investigation. Another interesting and important future direction is to incorporate a partial breadth-first search mechanism into  the current PPO algorithm  to increase the success rate and reduce the total length of the output sequences,  at the cost of a slightly more time and memory consuming training process. 
In addition, our proposed PPO algorithm may carry over straightforwardly to other scenarios, including quantum control problems \cite{Dong2010Quantum} and digital quantum simulations \cite{Suzuki1991General,Childs2021Theory,Berry2015Simulating,Low2019Hamiltonian,Bolens2021Reinforcement} for both closed and open systems.



In summary, we have rigorously proved that, in sharp contrast to the case of unitary compiling,  it is impossible to compile an arbitrary channel to arbitrary accuracy with any given finite elementary channel set, regardless of the length of the decomposition sequence.  We analytically constructed a general scheme to decompose an arbitrary channel into a sequence of elementary channels plus a unitary gate, hence reducing the task of channel compiling to unitary compiling. To reduce the count of certain experimentally expensive gates, we further introduced a DRL algorithm that is generally applicable to any elementary gate set and uses no ancillary qubit. We benchmarked the performance of our algorithm with an example concerning quantum compiling of Majorana fermions, and demonstrated that our approach can  reduce the use of expensive gates efficiently and effectively. Our results shed new light on the general problem of quantum compiling, which provides a valuable guide for future studies in both theory and experiment. 



\textit{Acknowledgement}.---We thank Zhengwei Liu, Weikang Li, Yuanhang Zhang, Qi Ye, Xun Gao, and Dong Yuan for helpful discussions. This work was supported by the start-up fund from Tsinghua University (Grant No. 53330300320), the National Natural Science Foundation of China (Grant. No. 12075128), and the Shanghai Qi Zhi Institute.

\bibliographystyle{apsrev4-1-title}
\bibliography{TopoChanCompRLTex}
\clearpage

\makeatletter
\setcounter{figure}{0}
\setcounter{equation}{0}
\renewcommand{\thefigure}{S\@arabic\c@figure}
\renewcommand \thetable{S\@arabic\c@table}

\begin{center} 
    {\large \bf Supplementary Materials for: Weighted Quantum Channel Compiling through Proximal Policy Optimization}
\end{center} 

In the Supplementary Materials, we provide more details about the proofs for the two theorems in the main text. We also provide supplementary notes on deep reinforcement learning, proximal policy optimization (PPO) algorithms, and a detailed description of our algorithm with more numerical results. We additionally provide a brief recapitulation for encoding and braiding Majorana fermions.

\section*{A. Proof for the Theorem 1}
We recall Eq. (1) in the main text which provides a matrix-form mapping representation for a linear CPTP quantum channel $\EE$ as 
\begin{equation}\label{suppeq:chan_matrix}\tag{A1}
\EE\to\mathcal T=\begin{pmatrix}1 & 0\\t & T\end{pmatrix}:\frac{1+\bm{a}\cdot\bm{\sigma}}{2}\to\frac{1+\bm{a}'\cdot\bm{\sigma}}{2},
\end{equation}
where $T$ is called a distortion matrix, $t$ is called a center shift and the state vectors $\bm a'$ and $\bm a$ are chosen within the Bloch sphere satisfying the relation $\bm a'=T\bm a+t$. This representation indicates that the channel $\EE$ maps the Bloch sphere into an ellipsoid. To guarantee the physical feasibility, the ellipsoid must be enveloped within the original Bloch sphere and $\bm |a|^2,\bm |a'|^2\leq 1$. Therefore, $T$ cannot have an eigenvalue that has magnitude larger than $1$. Moreover, if all eigenvalues of $T$ has magnitude $1$, then $t=0$ and $\EE$ is a unitary gate. For simplicity, we represent a quantum channel $\EE$ with distortion matrix $T$ and center shift $t$ as $\EE(T,t)$. 


As mentioned in the main text, the distance measure between channels used in this paper is the Schatten one-norm \cite{Kliesch2011Dissipative}, which measures the maximal $1$-norm distance between the output states of different quantum channels with the same quantum state $\rho$ chosen from $\HH_S$ as the input state. For single-qubit states $\rho_1=(1+\bm a_1\cdot\bm \sigma)/2$ and $\rho_2=(1+\bm a_2\cdot\bm \sigma)/2$, the trace distance between them reads $||\rho_1-\rho_2||_1=|\bm a_1-\bm a_2|/2$. This indicates that the distance between channels $\EE_1(T_1,t_1)$ and $\EE_2(T_2,t_2)$ is $D$ if $\max_{\rho=(1+\bm a\cdot\bm\sigma)/2\in\HH_S}|(T_1-T_2)\bm a+t_1-t_2|=2D$.

Now we start the proof for the first part of the Theorem 1. We first introduce the following lemma.

\textit{Lemma 1. }Suppose we have two single-qubit quantum channels $\EE_1(T_1,t_1)$ and $\EE_2(T_2,t_2)$,and $|\det(T_1)|>|\det(T_2)|$. If $|\det(T_1)|-|\det(T_2)|>6\epsilon$ where $\epsilon$ is a constant strictly smaller than the magnitude of any eigenvalue of $T_1$, then $||\EE_1-\EE_2||_{1\to1}>\epsilon$.

\begin{proof}
$\EE_1(T_1,t_1)$ and $\EE_2(T_2,t_2)$ in Eq.~\eqref{suppeq:chan_matrix} map the Bloch sphere into two ellipsoids. Consider quantum channels $\EE_1(T_1,t_1)$, $\EE_2'(T_2,t_1)$, and $\EE_2(T_2,t_2)$, and suppose $\rho^*=\frac{1+\bm a\cdot\bm\sigma}{2}=\arg\max_{\rho\in\HH_S}||\EE_1(\rho)-\EE_2'(\rho)||_1$. Notice that $\EE_1$ and $\EE_2'$ have the same center shift $t_1$, we can find another quantum state that also yields the maximal output state distance as $\rho^*_{\text{opp}}=\frac{1-\bm a\cdot\bm\sigma}{2}=\arg\max_{\rho\in\HH_S}||\EE_1(\rho)-\EE_2'(\rho)||_1$ according to the symmetry property of the ellipsoid. Since the ellipsoid derived by $\EE_2$ could be regarded as the ellipsoid derived by $\EE_2'$ with an additional center shift $t_2-t_1$, at least one of the distances $||\EE_1(\rho^*)-\EE_2(\rho^*)||_1$ and $||\EE_1(\rho^*_{\text{opp}})-\EE_2(\rho^*_{\text{opp}})||_1$ is not smaller than $||\EE_1(\rho^*)-\EE_2'(\rho^*)||_1$. Therefore, we conclude that $||\EE_1-\EE_2||_{1\to 1}\geq||\EE_1-\EE_2'||_{1\to1}$.

Now we prove that if $|\det(T_1)|-|\det(T_2)|> 6\epsilon$, then $||\EE_1-\EE_2||_{1\to 1}\geq||\EE_1-\EE_2'||_{1\to1}>\epsilon$. We denote the eigenvalues of $T_1$ as $\lambda_1$, $\lambda_2$, and $\lambda_3$ with $\lambda_1\lambda_2\lambda_3=\det(T_1)$, and the eigenvalues of $T_2$ as $\lambda_1'$, $\lambda_2'$, and $\lambda_3'$ with $\lambda_1'\lambda_2'\lambda_3'=\det(T_2)$. If $||\EE_1-\EE_2'||\leq\epsilon$, the ellipsoid produced by $\EE_2'$ would envelope all quantum states that are $2\epsilon$ distance within the ellipsoid produced by $\EE_1$. Therefore, the ellipsoid with semi-major axis of length $|\lambda_1|-2\epsilon,|\lambda_2|-2\epsilon,|\lambda_3|-2\epsilon$ and the same direction with the ellipsoid produced by map $\EE_1$ should fall completely within the ellipsoid produced by $\EE_2'$. That is to say, $|\det(T_2)|=|\lambda_1'||\lambda_2'||\lambda_3'|\geq(|\lambda_1|-2\epsilon)(|\lambda_2|-2\epsilon)(|\lambda_3|-2\epsilon)$. As we mentioned before, $T_1$ and $T_2$ cannot contain eigenvalues with absolute value larger than $1$, thus $|\det(T_2)|\geq|\det(T_1)|-6\epsilon$. This completes the proof for the lemma. 
\end{proof}

Then we prove the first part of the Theorem 1 based on the above lemma. Suppose we have finite number $N$ of elementary channels $\EE_1(T_1,t_1),...,\EE_N(T_N,t_N)$ and arbitrary unitary gates in the channel compiler and the distortion matrix $T_i$ has determinant of absolute value $d_i=|\det(T_i)|$. Without loss of generality, we further assume that $1> d_1\geq d_2\geq...\geq d_N\geq0$. Therefore, any channel $\EE(T,t)$ that can be represented by a sequence of elementary channels and unitary gates satisfies $\det(T)\leq d_1$, or $\det(T)=1$ when the sequence only consists unitary gates.

Considering the compilation of a target channel $\EE^*(T^*,t^*)$ with $\det(T^*)=\frac{1+d_1}{2}$ and accuracy demand $\epsilon<\frac{1-d_1}{12}$, any decomposition sequence, whose corresponding generated channel is denoted  as $\EE(T,t)$, satisfies either $\det(T)\leq d_1<\det(T^*)-6\epsilon$ or $\det(T)=1>\det(T^*)+6\epsilon$. Therefore, according to the Lemma $1$ this target channel cannot be decomposed into elementary channels and unitary gates under such accuracy demand $\epsilon$. This completes the proof for part $(1)$ of the Theorem 1 in the main text.


For part $(2)$ of this theorem, we assume that the target channel is $\EE^*(T^*,t^*)$ and $T^*$ has eigenvalues $\lambda_1^*$, $\lambda_2^*$, and $\lambda_3^*$ with $|\lambda_1^*|\geq|\lambda_2^*|\geq|\lambda_3^*|$ and orthonormal eigenvectors $v_1^*$, $v_2^*$ and $v_3^*$. We denote $t^*=(t_1^*,t_2^*,t_3^*)$ under basis $\{v_1^*,v_2^*,v_3^*\}$. Without loss of generality, we assume $t_i^*>0$ and suppose the accuracy demand is $\epsilon$. Now, we propose a four-step procedure to decompose the target channel into a sequence of unitary gates and channels chosen from $14$ elementary channels. 

\textit{Step 1.} We consider realizing a channel $\EE_{\text{step 1}}(T_\text{step 1},t_\text{step 1})$ where $T_\text{step 1}=\text{diag}\{|\lambda_1^*|,|\lambda_2^*|,|\lambda_3^*|\}$ and $t_\text{step 1}=(t_1^*,t_2^*,t_3^*)$. We construct the following $14$ elementary channels $\EE_1(T_1,t_1),...,\EE_{14}(T_{14},t_{14})$ using a parameter $\delta(\epsilon)$ to be fixed later:
\begin{subequations}
\begin{equation}\label{suppeq:thm1_step1_1}\tag{A2}
T_1=...=T_8=\text{diag}\{1-\delta,1-\delta,1-\delta\},
\end{equation}
\begin{equation}\label{suppeq:thm1_step1_2}\tag{A3}
T_9=...=T_{12}=\text{diag}\{1,1-\delta,1-\delta\},
\end{equation}
\begin{equation}\label{suppeq:thm1_step1_3}\tag{A4}
T_{13}=T_{14}=\text{diag}\{1,1,1-\delta\},
\end{equation}
\begin{equation}\label{suppeq:thm1_step1_4}\tag{A5}
t_1=t_9=t_{13}=\bm 0, t_8=(\delta,\delta,\delta)^T,t_2=(\delta,0,0)^T,
\end{equation}
\begin{equation}\label{suppeq:thm1_step1_5}\tag{A6}
t_3=t_{10}=(0,\delta,0)^T,t_4=t_{11}=t_{14}=(0,0,\delta)^T,
\end{equation}
\begin{equation}\label{suppeq:thm1_step1_6}\tag{A7}
t_5=(\delta,\delta,0)^T,t_6=(\delta,0,\delta)^T,t_7=t_{12}=(0,\delta,\delta)^T.
\end{equation}
\end{subequations}
Denoting $k_i=\lceil\min\{\log_{(1-\delta)}|\lambda_i^*|,\log_{(1-\delta)}\delta\}\rceil,i=1,2,3$, we have $||\lambda_i^*|-(1-\delta)^{k_i}|<\delta,i=1,2,3$. We introduce a procedure to use the above elementary channels to compile $T_\text{step 1}$ within distance $6\delta$ using a sequence of $k_3$ elementary channels. We exploit the Table. \ref{tab:supp1} to record each elementary channel in this sequence. In this sequence, we introduce three $\{0,1\}$ strings $s_x=s_{x,1}...s_{x,k_1}$, $s_y=s_{y,1}...s_{y,k_2}$, and $s_z=s_{z,1}...s_{z,k_3}$. For $s_{i,j}$ ($i=x,y,z$), it is $0$ or $1$ when the center shift of the $j$-th channel in the sequence on $i$-axis is $0$ or $\delta$. We use $T^{ii}_j$ ($i=1,2,3, j=1,...,k_3$) to represent the $(i,i)$-th entry of the distortion matrix $T_j$ for the $j$-th channel in the sequence.

\begin{table}
\centering
\begin{tabular}{|r|r|r|r|r|r|r|r|r|r|}  
\hline  
Seq. Pos. & $1$ &...& $k_1$ & $k_1+1$ & ... & $k_2$ & $k_2+1$ & ... & $k_3$\\
\hline
$T^{11}_j$ & $1-\delta$ & ... & $1-\delta$ & $1$ & ... & $1$ & $1$ & ... & $1$\\  
\hline
$s_x$ & $s_{x,1}$ & ... & $s_{x,k_1}$ & - & - & - & - & - & - \\
\hline
$T^{22}_j$& $1-\delta$ & ... & $1-\delta$ & $1-\delta$ & ... & $1-\delta$ & $1$ & ... & $1$\\  
\hline
$s_y$ & $s_{y,1}$ & ... & $s_{y,k_1}$ & $s_{y,k_1+1}$ & ... & $s_{y,k_2}$ & - & - & - \\
\hline
$T^{33}_j$& $1-\delta$ & ... & $1-\delta$ & $1-\delta$ & ... & $1-\delta$ & $1-\delta$ & ... & $1-\delta$\\  
\hline
$s_z$ & $s_{z,1}$ & ... & $s_{z,k_1}$ & $s_{z,k_1+1}$ & ... & $s_{z,k_2}$ & $s_{z,k_2+1}$ & ... & $s_{z,k_3}$ \\
\hline
\end{tabular}
\caption{A table illustration for our sequence of length $k_3$ to approximate the channel $\EE_\text{step 1}$ within distance $6\delta$ consisting of elementary channels in Eqs. \eqref{suppeq:thm1_step1_1}-\eqref{suppeq:thm1_step1_6}. The first row shows the position of the elementary channel in the decomposition sequence (abbreviated as Seq. Pos.). 
For the $j$-th elementary channel in this sequence, it has a distortion matrix $T_j=\text{diag}\{T_j^{11},T_j^{22},T_j^{33}\}$ and a center shift $t_j=(s_{x,j}\delta,s_{y,j}\delta,s_{z,j}\delta)$ for $j=1,...,k_1$, $t_j=(0,s_{y,j}\delta,s_{z,j}\delta)$ for $j=k_1+1,...,k_2$, and $t_j=(0,0,s_{z,j}\delta)$ for $j=k_2+1,...,k_3$.}
\label{tab:supp1}
\end{table}

The sequence in the above table composes a channel $\EE'(T',t')$, where $T'=\text{diag}\{(1-\delta)^{k_1},(1-\delta)^{k_2},(1-\delta)^{k_3}\}$ and $t'=(t_1',t_2',t_3')$ with each element
\begin{subequations}
\begin{equation}\label{suppeq:thm1_step1_t1}\tag{A8}
t_1'=\delta\sum_{j=1}^{k_1}s_{x,j}(1-\delta)^{j-1},
\end{equation}
\begin{equation}\label{suppeq:thm1_step1_t2}\tag{A9}
t_2'=\delta\sum_{j=1}^{k_2}s_{y,j}(1-\delta)^{j-1},
\end{equation}
\begin{equation}\label{suppeq:thm1_step1_t3}\tag{A10}
t_3'=\delta\sum_{j=1}^{k_3}s_{z,j}(1-\delta)^{j-1}.
\end{equation}
\end{subequations}

From Eqs.~\eqref{suppeq:thm1_step1_t1}-\eqref{suppeq:thm1_step1_t3}, as $||\lambda_i^*|-(1-\delta)^{k_i}|<\delta,i=1,2,3$, we can observe that by changing $\{0,1\}$ strings $s_x,s_y,s_z$, $t_1',t_2',t_3'$ forms a $\delta$-net of interval $[0,1-|\lambda_1^*|]$, $[0,1-|\lambda_2^*|]$, and $[0,1-|\lambda_3^*|]$, which include all possible value of $t_1^*,t_2^*,t_3^*$ because the output ellipsoid of the quantum channel should be within the original Bloch sphere. Given a center shift element $t_i^*\in[0,1-|\lambda_i^*|]$, we could calculate $s_i$ by extending $t_i^*$ into the summation over a series $\delta\sum_{j=1}^{k_1}s_{i,j}(1-\delta)^{j-1}$ on $(1-\delta)$.Therefore, for an arbitrary $\EE_\text{step 1}$, we can find strings $s_x$, $s_y$, and $s_z$ such that each entry of $T'$ and $t'$ differs from $T_\text{step 1}$ and $t_\text{step 1}$ by $\delta$ at most. Hence, the total distance is strictly smaller than the sum of the distance, which is $6\delta$.

\textit{Step 2. }If eigenvalues $\lambda_1^*,\lambda_2^*,\lambda_3^*$ contain complex values, i.e. $\lambda_2^*=re^{i\theta},\lambda_3^*=re^{-i\theta},r\in\RR$, we use a unitary gate that can be represented as an affine map $\EE_R(T_R,t_R)$ with $t_R=\bm 0$ and
\begin{equation}\label{suppeq:thm1_step2}\tag{A12}
T_R=\begin{pmatrix}1 & 0 & 0\\ 0 & \cos\theta & \sin\theta\\ 0 & -\sin\theta & \cos\theta\end{pmatrix}.
\end{equation}
Such a unitary gate can introduce the complex phase $e^{i\theta}$ and $e^{-i\theta}$ for $\lambda_2^*$ and $\lambda_3^*$, respectively. Therefore, in the following steps, we only need to consider the case where all eigenvalues are real.

\textit{Step 3. }If the $T^*$ contains real negative eigenvalues, we consider the following three unitary transformations with affine map representations $\EE_X(T_X,t_X)$, $\EE_Y(T_Y,t_Y)$, and $\EE_Z(T_Z,t_Z)$ with $t_X=t_Y=t_Z=\bm 0$ and
\begin{equation}\label{suppeq:thm1_step3}\tag{A13}
T_X=\begin{pmatrix} -1 & 0 & 0\\ 0 & 1 & 0\\0 & 0 & 1 \end{pmatrix},T_Y=\begin{pmatrix} 1 & 0 & 0\\ 0 & -1 & 0\\0 & 0 & 1 \end{pmatrix},T_Z=\begin{pmatrix} 1 & 0 & 0\\ 0 & 1 & 0\\0 & 0 & -1 \end{pmatrix}.
\end{equation}
Since their products are still unitary transformations, we need only one unitary transformation to turn positive eigenvalues obtained in step $1$ into negative values. 

\textit{Step 4. }In this step we apply a unitary transformation to transfer the current basis to the $\{v_1^*,v_2^*,v_3^*\}$ basis.

In the above four steps, the unitary gates in step $2$ to $4$ could be combined as one unitary gate, and according to Ref. \cite{Nielsen2010Quantum} this unitary gate can be approximated within error $\delta$ by a sequence of elementary gates chosen from a universal gate set. Therefore, the total error cannot exceed the sum of error in all steps, which is $7\delta$. By fixing $\delta=\frac{\epsilon}{7}$, we can decompose an arbitrary quantum channel into a sequence of unitary gates and elementary channels chosen from the $14$ elementary channels $\EE_1,...,\EE_{14}$ constructed in Eqs.~\eqref{suppeq:thm1_step1_1}-\eqref{suppeq:thm1_step1_6}. 

Following the four steps above, we can also bound the length of the sequence for the compilation. In step $2$-$4$, we need one unitary gate while in step $1$, the length of the table does not exceed $\log_{(1-\delta)}\delta+1$. In practice, $\delta=\frac{\epsilon}{7}$ is usually a small number close to $0$. Therefore, we can do the approximation $\log_{(1-\delta)}\delta=\frac{\ln(\delta)}{\ln(1-\delta)}\approx\frac{1}{\delta}\ln(\frac{1}{\delta})=O(\frac{1}{\epsilon}\log(\frac1\epsilon))$. This indicates that the length of the entire sequence is $O(\frac{1}{\epsilon}\log(\frac1\epsilon))$. This completes the proof of the part $(2)$ of the Theorem 1 in the main text.


We can extend the Theorem 1 to the multi-qubit case. As mentioned in the main text, for a $d$-dimensional quantum state $\rho\in\OO(\HH_S)$, a canonical and orthonormal basis \cite{Wang2015Algorithmic,Bruning2012Parametrizations} $\{O_\alpha\},O_\alpha\in\OO(\HH_S)$ satisfies (i) $O_0=I$, (ii) $\text{tr}(O_\alpha)=0,\forall\alpha\neq0$, and (iii) $\text{tr}(O_\alpha^\dagger O_\beta)=\delta_{\alpha\beta}$. An arbitrary density operator $\rho$ can be written as $\rho=\frac 1d(I+\sum_{\alpha=1}^{d^2-1}p_\alpha Q_\alpha)$, where $Q_\alpha=\sqrt{d(d-1)}O_\alpha$. The parameters in $\{p_\alpha\}$ form the polarization vector $\bm{p}=(p_1,...,p_{d^2-1})$ of a $(d^2-1)$-dimensional ball with $||p||_2=1$ representing pure states and $||p||_2<1$ representing mixed states. Since a quantum state $\rho$ can be represented as a vector within a ball, a quantum channel $\EE:\OO(\HH_S)\to\OO(\HH_S)$ can be written as an affine map represented by a distortion matrix $T\in\RR^{(d^2-1)\times(d^2-1)}$ and a center shift $t\in\RR^{d^2-1}$
\begin{subequations}
\begin{equation}\label{suppeq:coro1_map1}\tag{A14}
\EE\to\mathcal T=\begin{pmatrix}1 & 0\\t & T \end{pmatrix},\mathcal{T}_{ij}=\frac1d\text{tr}[Q_\alpha\EE(Q_\beta)],
\end{equation}
\begin{equation}\label{suppeq:coro1_map2}\tag{A15}
\bm p\to T\bm p+t.
\end{equation}
\end{subequations}
Notice that the above affine map is similar to the single-qubit case mentioned in the main text, we can extend part (1) of the Theorem 1 straightforwardly to the multi-qubit scenario.  

For the second part, we assume the target channel to be $\EE^*(T^*,t^*)$ and $T^*$ has eigenvalues $\lambda_1^*,...,\lambda_{d^2-1^*}$. Without loss of generality, we can rank $\lambda_i^*$ in decreasing order of magnitude such that $|\lambda_1^*|\geq...\geq|\lambda_{d^2-1}^*|$. Inherited from the proof for single-qubit channel, $k_i$ is defined to be $k_i=\lceil\min\{\log_{(1-\delta)}|\lambda_i^*|,\log_{(1-\delta)}\delta\}\rceil,i=1,...,d^2-1$. If we directly use the constructive approach in the proof for the Theorem 1, we will create $2^{d^2}-2$ elementary channels $\EE_1(T_1,t_1),...,\EE_{2^{d^2}-2}(T_{2^{d^2}-2},t_{2^{d^2}-2})$. The first $2^{d^2-1}$ channels $\EE_1(T_1,t_1),...,\EE_{2^{d^2-1}}(T_{2^{d^2-1}},t_{2^{d^2-1}})$ have the same distortion matrix $T_1=...=T_{2^{d^2-1}}=\text{diag}\{1-\delta,...,1-\delta\}$ and their center shifts go through $2^{d^2-1}$ cases in which each element of the center shift can be $0$ or $\delta$. The next $2^{d^2-2}$ elementary channels that follows are $\EE_{2^{d^2-1}+1}(T_{2^{d^2-1}+1},t_{2^{d^2-1}+1}),...,\EE_{2^{d^2-1}+2^{d^2-2}}(T_{2^{d^2-1}+2^{d^2-2}},\\t_{2^{d^2-1}+2^{d^2-2}})$. They share the same distortion matrix $\text{diag}\{1, 1-\delta,...,1-\delta\}$ and their center shifts go through all cases that have the first element to be $0$ and other elements to be either $0$ or $\delta$. The rest channels are constructed similarly until the last two channels $\EE_{2^{d^2}-3}(T_{2^{d^2}-3},t_{2^{d^2}-3}),\EE_{2^{d^2}-2}(T_{2^{d^2}-2},t_{2^{d^2}-2})$, which have the distortion matrix $\text{diag}\{1,...,1,1-\delta\}$ and center shifts $(0,....,0)$ and $(0,0,...,0,\delta)$. Similar to the previous proof, we hold strings $s_i,i=1,...,d^2-1$ with the $j$-th element $s_{i,j}=0,1$ representing whether the $i$-th element of the center shift for the $j$-th channel of the sequence is $0$ or $\delta$. 

Under this construction, the error in total can be bounded above by $(2(d^2-1)+1)\delta$. Therefore, we fix $\delta=\frac{\epsilon}{2(d^2)-1}$ to guarantee that the distance between our approximation and target channel is no more than $\epsilon$. The length of the sequence is still bounded by $O(\frac1\epsilon\log(\frac1\epsilon))$.

However, notice that the above construction requires $O(2^{d^2})$ elementary channels, which is double exponential to the number of qubits $n$. Here, we propose another construction that only requires $O(d^2)$ elementary channels. We still exploit the $4$-step compiling process in the previous section. Step $2$ to $4$ remain the same with previous construction to implement complex, negative eigenvalues in the distortion matrix and perform orthonormal basis transformation, In step $1$, we simply use $\EE_1(T_1,t_1),...,\EE_{2(d^2-1)}(T_{2(d^2-1)},t_{2(d^2-1)})$ with $T_{2i-1}=T_{2i}=\text{diag}\{1,...,1,1-\delta,1,...,1\}$ where the $i$-th diagonal element is $1-\delta$, $t_{2i-1}=(0,...,0)$, and $t_{2i}=(0,...,0,\delta,0,...,0)$ with the $i$-th element being $\delta$.

Under this construction, step $1$ could be decomposed into sub-steps compiling $\EE_{\text{step i}}(T_{\text{step i}},t_{\text{step i}})$ with $T_{\text{step i}}=\text{diag}\{1,...,1,|\lambda_i^*|,1, ...,1\},t_{\text{step i}}=(0,...,0,t_i^*,0,..,0)$ using $\EE_{2i-1},\EE_{2i}$ separately, where $|\lambda_i^*|$ and $t_i$ are the $i$-th diagonal element of distortion matrix and the $i$-th element of the center shift. In the $i$-th sub-step, we keep a $k_i$ length $0$-$1$ string $s_i$ and decompose $\EE_{\text{step i}}$ into a sequence containing $k_i$ elementary channels. If the $j$-th element $s_{i,j}$ of $s_i$ is $0$, we add $\EE_{2i-1}$ to the sequence and otherwise we add $\EE_{2i}$. Therefore, the $i$-th element of approximation for the center shift is $\sum_{j=1}^{k_i}s_{i,j}\delta(1-\delta)^{j-1}$, which form a $\delta$-net in range $[0,1-|\lambda_i^*|]$ as $||\lambda_i^*|-(1-\delta)^{k_i}|<\delta,i=1,...,d^2-1$. Therefore, in each sub-step we can approximate $|\lambda_i^*|$ and $t_i^*$ within distance $\delta$. 

The total distance in this step cannot exceed $2(d^2-1)\delta$, which is the same with the previous construction. We can still fix $\delta=\frac{\epsilon}{2(d^2)-1}$. It is worthwhile to mention that there exists a trade-off between the above two constructions. Though the second construction only requires $O(d^2)$ elementary channels, the output sequence would have a total length $O(d^2\frac1\epsilon\log(\frac1\epsilon))$, which is exponential in the number of qubits $n$.

\section*{B. Proof for the Theorem 2} 
In this section, we provide the detailed proof for the Theorem 2 in the main text. To derive a lower bound, we exploit the volume method \cite{Kitaev2002Classical,Harrow2002Efficient} based on the constraint that the whole space of $SU(d)$ should be covered by the $\epsilon$-balls centered by the gates that could be implemented by an elementary gate sequence.

We start with the case when the subgroup $G=\langle g_1,...,g_n\rangle$ generated by $\{g_1,...,g_n\}$ is finite. We denote the size of $G$ as $|G|=K$. Consider compilation with fewer than $t$ $g^*$ gates and unlimited number of gates chosen from $G$. If no $g^*$ gate is used, we can only compile the $K$ gates in $G$. When we use at least one $g^*$ gates, any sequence containing $0<k\leq t$ $g^*$ gates could always be written as $g=(g_{i_{11}}g^*g_{i_{12}})...(g_{i_{k1}}g^*g_{i_{k2}})$ where $g_{i_{11}},g_{i_{12}},...,g_{i_{k1}},g_{i_{k2}}$ are chosen from $G$. Consider the subset $G^*=\{g_sg^*g_t|g_s,g_t\in G\}$ that can generate a dense subgroup of $SU(d)$. Any sequence that contains $k$ $g^*$ gates can be regarded as $k$ gates in $G^*$. Therefore, the number of $g^*$ gates can be regard as the number of gates from $G^*$. As the size of $G^*$ is $|G^*|=K^2$, the possible number of gate sequences containing no more than $t$ $g^*$ could be bounded above by $|G^*|+|G^*|^2+...+|G^*|^{t}\leq O(K^{2t+2})$. Therefore, the total number of gates we can accurately compile with a gate sequence with fewer than $t$ $g^*$ gates is bounded above by $O(K^{2t+2})$. 

We exploit the normalized Haar measure on $SU(d)$ space such that the volume of $SU(d)$ is one \cite{Harrow2002Efficient}. Under this measure, the volume of $\epsilon$-ball in $SU(d)$ group scales as $\Theta(\epsilon^{d^2-1})$. If any gate in $SU(d)$ can be approximated within distance $\epsilon$, all $\epsilon$-balls centered by possible gates sequences generated by no more that $t$ $g^*$ gates should cover $SU(d)$. Hence, we can deduce that $k\geq\frac{d^2-1}{\log{|G|}}\log(\frac1\epsilon)$. This completes the proof of the Theorem 2 in the main text for a finite $G$.

When the subgroup $G$ is infinite, the lower bound given by the volume method would reduce to $\Omega(1)$. This lower bound, however, is not as trivial as it seems. Indeed, we could even find a simple extreme example, in which one can compile arbitrary gates with only two $g^*$ gates and unlimited number of gates chosen from an infinite $G$. Consider a single-qubit gate compilation with $g^*=H$ and $G=\langle R_x(\alpha\pi)\rangle$ with a generator $R_x(\alpha\pi)$ of an irrational number $\alpha$. In this example, $G$ is an infinite group which could approximate all the rotations along x-axis within arbitrary accuracy demands. Notice that an arbitrary single-qubit gate could be written as $R_x(\theta_1)R_z(\theta_2)R_x(\theta_3)$ \cite{Nielsen2010Quantum} and $R_z(\theta_2)=HR_x(\theta_2)H$, any single-qubit gate could be compiled by gates chosen from $G$ and at most two $H$ gates. 

In the main text, we focus on the quantum compiling on Majorana fermions and non-topological $T$ gates. Based on this compilation scheme, various algorithms have been proposed to reduce the $T$ count for both multi-qubit unitary compilation \cite{Gheorghiu2021TCount,Heyfron2018Efficient} and single-qubit unitary compilation \cite{Selinger2015Efficient,Matsumoto2008Representation}. However, these algorithms focus on providing an optimal number of $T$ gates for a given unitary. Compare with these approaches, our result provide a lower bound for the worst case overall unitaries chosen from $SU(d)$.

\section*{C. Deep reinforcement learning and PPO algorithm}
In this section, we give a brief introduction to the deep reinforcement learning (DRL) and the PPO algorithm we exploit to compile unitary gates. 

\subsection*{I. Introduction to Reinforcement Learning and Policy Gradient}

To formalize the learning process, we introduce some basic concepts and notations. A reinforcement learning (RL) aims to train a decision-making agent in a Markovian decision process. This process involves a state set $S$ and an action set $A$. In step $n$, the agent chooses an action $a_n\in A$ while the environment shifts from $s_n\in S$ to $s_{n+1}\in S$, providing the agent a scalar value reward $r_n$ as feedback. In a Markovian decision process, the new state $s_{n+1}$ and reward $r_n$ only depend on the former state $s_n$ and action $a_n$. Therefore, the RL process can be written as a trajectory $\tau:s_0\to a_0\to r_0\to s_1\to ...\to a_{N-1}\to r_{N-1}\to s_{N}$, where $N$ is the maximal number of steps in an iteration. 
 
The core problem in RL is to learn a policy to choose the optimal action $a_n$ given the environment state $s_n$. Therefore, a policy function $\pi$ is introduced to map the states to the action probability distributions. The agent uses this function $\pi$ to perform decision making tasks and choose the action $a\sim\pi(s)$ according to the policy for the state $s$. In deep reinforcement learning, a policy $\pi_\theta$ is represented by a policy network with parameters $\theta$. To evaluate the performance of $\pi_\theta$, an objective function  is defined to be the expected return over all complete trajectories:
\begin{equation}\label{suppeq:J}\tag{C1}
J(\pi_\theta)=\mathbb{E}_{\tau\sim\pi_\theta}[\sum_{n=1}^N\gamma^nr_n],
\end{equation}
where $\gamma$ is the discount factor and the expectation is obtained over all trajectories sampled from $\pi_\theta$. 

To find an optimal policy, we have to find the policy that maximize the objective function. A straight forward approach is to perform gradient descent on policy to solve the optimization.
\begin{equation}\label{suppeq:PolicyGDStep}\tag{C2}
\theta\leftarrow\theta+\alpha\nabla_{\theta}J(\pi_\theta),
\end{equation}
where $\alpha$ is a scalar factor known as the learning rate and $\nabla_\theta J(\pi_\theta)$ is known as the \textit{policy gradient}. It is proved that the policy gradient could be calculated by \cite{Graesser2019Foundations}:
\begin{equation}\label{suppeq:PolicyG}\tag{C3}
\nabla_{\theta}J(\pi_\theta)=\mathbb{E}_{\tau\sim\pi_\theta}\left[\sum_{n=0}^NR_n(\tau)\nabla_\theta\log\pi_\theta(a_n|s_n)\right],
\end{equation}
where the action $a_n\sim\pi_\theta$ is sampled from probability distribution $\pi_\theta(a_n|s_n)$ given by the policy at step $n$ and $R_n(\tau)=\sum_{n'= n}^N\gamma^{n'-n}r_n'$ is the discounted sum of reward from current step $n$ to the end of the trajectory. The algorithm for this simple policy gradient method is summarized in Algorithm \ref{pseudo_PG}

\begin{algorithm}
\caption{The Policy Gradient Algorithm}
\begin{algorithmic}[1]  
\Require The environment $\mathcal{E}$, the initial policy parameters $\theta_0$, the number of epochs $T$, the step size $\alpha$.
\Ensure The optimal policy $\pi_{\theta^*}$
\For {$i=0, 1,2, \dots, T$}
\State Running current policy $\pi_{\theta_{i}}$ in $\mathcal{E}$ and obtain a set of trajectories $\mathcal{D}_i=\{\tau_i\}$.
\State{Compute the policy gradient:
\[(\nabla_{\theta}J)_i=\frac{1}{|\mathcal{D}_i|}\sum_{\tau \in \mathcal{D}_i}R(\tau)\sum_{n=0}^{N(\tau)}\nabla_\theta\log\pi_{\theta_i}(a_n|s_n),\]}
\State{Update the policy by gradient ascent:
\[\theta_{i+1}=\theta_i+\alpha (\nabla_{\theta}J)_i\]}
\EndFor\\
\Return $\pi_{\theta^*}=\pi_{\theta_{T+1}}$.
\end{algorithmic}  \label{pseudo_PG}
\end{algorithm}

However, in practical policy gradient the parameter space and the policy space do not always map congruently. This fact makes it challenging to find a step size $\alpha$. If $\alpha$ is chosen as a small constant, more iterations is potentially required in the training process. If $\alpha$ is chosen bigger, the agent would be vulnerable to a \textit{performance collapse} in which the agent chooses a bad action, resulting in a sudden drop in its performance. In addition, another issue of policy gradient descent is that it is \textit{sample-inefficient} because it does not reuse data. To address these two issues, the proximal policy optimization algorithms \cite{Schulman2017Proximal} are proposed.


\subsection*{II. Proximal Policy Optimization (PPO)}
Our algorithm employs proximal policy optimization (PPO) \cite{Schulman2017Proximal}, which is a policy gradient algorithm developed by \textsc{OpenAI}. The PPO is motivated by a algorithm called Trust Region Policy Optimization (TRPO) \cite{Schulman2015Trust}, which aims to find an optimal policy iteratively to maximize the $J(\pi)$ without causing a performance collapse. While TRPO applies second-order methods which are complex to compute, PPO uses first-order methods and tricks to keep the updated policy not changed too fast. PPO is much simpler to implement than TRPO, while performing well in practice.

To introduce PPO algorithm, We first define a value function $V_\pi(s)=\mathbb{E}_{s_0=s,\tau\sim\pi}[\sum_{n=1}^N\gamma^nr_n]$ and a value-action function $Q_\pi(s,a)=\mathbb{E}_{s_0=s,a_0=a,\tau\sim\pi}[\sum_{n=1}^N\gamma^nr_n]$ given state $s$, action $a$, discount factor $\gamma$ and policy $\pi$. These two functions are used to evaluate a state and a given state-action pair. The objective function used in the learning procedure is defined to be the expected reward of policy $\pi$ as $J(\pi)=\mathbb{E}_{\tau\sim\pi}[\sum_{n=1}^N\gamma^nr_n].$
 In the policy descent procedure, we will get another policy $\pi'$ in the next iteration given a current policy $\pi$. The objective function changes as:
\begin{equation}\label{suppeq:JimpCPI}\tag{C4}
J(\pi')-J(\pi)=\mathbb{E}_{\tau\sim\pi'}[\sum_{n=1}^N\gamma^nA_\pi(s_n,a_n)],
\end{equation}
where $A_\pi(s_n,a_n)=Q_\pi(s_n,a_n)-V_\pi(s_n)$ is defined as the advantage function. The relative policy performance $J(\pi')-J(\pi)$ provides a metric to measure the improvement of performance after a policy shift. Therefore, maximizing $J(\pi')$ is equivalent to maximizing $J(\pi')-J(\pi)$. 

To approximate Eq. \eqref{suppeq:JimpCPI}, we use the trajectories from the old policy $\tau\sim\pi$ and adjust with importance sampling weights $R_n(\pi)=\frac{\pi'(a_n|s_n)}{\pi(a_n,s_n)}$ \cite{Graesser2019Foundations}. This approximation is given as
\begin{equation}\label{suppeq:Jimp}\tag{C5}
J(\pi')-J(\pi)\approx J^{\text{CPI}}_\pi(\pi')=\mathbb{E}_{\tau\sim\pi'}[\sum_{n=1}^NA_\pi(s_n,a_n)R_n(\pi)],
\end{equation}
where $J^{\text{CPI}}_\pi(\pi')$ is known as a surrogate objective. The surrogate objective function can be additionally written as an average over both $\tau\sim\pi'$ and $n$ as $\mathbb{E}_{\tau\sim\pi',n}[A_\pi(s_n,a_n)R_n(\pi)]$. In our algorithm, we use an alternative version of PPO with clipped surrogate objective function
\begin{equation}\label{suppeq:Jimpclip}\tag{C6}
\begin{aligned}
J^{\text{CLIP}}=&\mathbb{E}_{\tau\sim\pi',n}[A_\pi(s_n,a_n)\times\\&\min\{R_n(\pi),\text{clip}(R_n(\pi),1-\epsilon,1+\epsilon)\}].
\end{aligned}
\end{equation}

The above equation is known as the clipped surrogate objective function and $\epsilon$ is a hyperparameter which defines the clipping bound $|R_n(\pi)-1|\leq\epsilon$. This parameter would decay during the training procedure. As the term $\text{clip}(R_n(\pi),1-\epsilon,1+\epsilon)A_\pi(s_n,a_n)$ bounds the value $J^{\text{CPI}}$, this objective function prevents the updates that create large and risky policy changes.

In objective $J^{\text{CLIP}}$, the most computationally costly parts are the weight $R_n(\pi)$ and advantage $A_\pi(s_n,a_n)$. However, these parts are required in any algorithm that optimizes the surrogate objective function. The rest calculations are essentially constant-time clippings and minimizings. Therefore, the clipped objective is relatively easy to compute and understand. The whole PPO algorithm is summarized in Algorithm \ref{pseudo_PPO}.

\begin{algorithm}
\caption{The  Proximal   Policy   Optimization Algorithm}
\begin{algorithmic}[1]  
\Require The environment $\mathcal{E}$, the initial policy parameters $\theta_0$, the initial value function parameters $\phi_0$, the number of epochs $T$, the step size $\alpha$.
\Ensure The optimal policy $\pi_{\theta^*}$
\For {$i=0, 1,2, \dots, T$}
\State Running current policy $\pi_{\theta_{i}}$ in $\mathcal{E}$ and obtain a set of trajectories $\mathcal{D}_i=\{\tau_i\}$.
\State Compute the advantage function $A_{\pi_i}(s,a)$ by using the estimation of value function $V_{\phi_i}$ 
\State {Compute the clipped surrogate objective:
\[J^{\text{CLIP}}_i=\sum_{\tau\in\mathcal{D}_i}\frac{\sum_{n=0}^{N(\tau)}\frac{A_\pi(s_n,a_n)}{N(\tau)}\min\{R_n(\pi),\text{clip}(R_n(\pi),\epsilon)\}}{|\mathcal{D}_i|}\]
}
\State{Update the policy by optimize the objective, this step is done by gradient ascent:
\[\theta_{i+1}=\text{argmax}_{\theta} J^{\text{CLIP}}_i\]}
\State{Update the estimation of value function by minimize the MSE, this step is done by gradient decent:
\[\phi_{i+1}=\text{argmax}_{\phi}\frac{1}{|\mathcal{D}_i|}\sum_{\tau\in\mathcal{D}_i}\sum_{n=0}^{N(\tau)}\frac{(V_{\phi_i}(s_n)-\sum_{t=n}^{N(\tau)}r_t)^2}{N(\tau)}\]}
\EndFor\\
\Return $\pi_{\theta^*}=\pi_{\theta_{T+1}}$.
\end{algorithmic}  \label{pseudo_PPO}
\end{algorithm}

\section*{D. Supplementary note on the algorithm and numerical experiments}
\begin{figure*}
    \centering
    \includegraphics[width=0.99\textwidth]{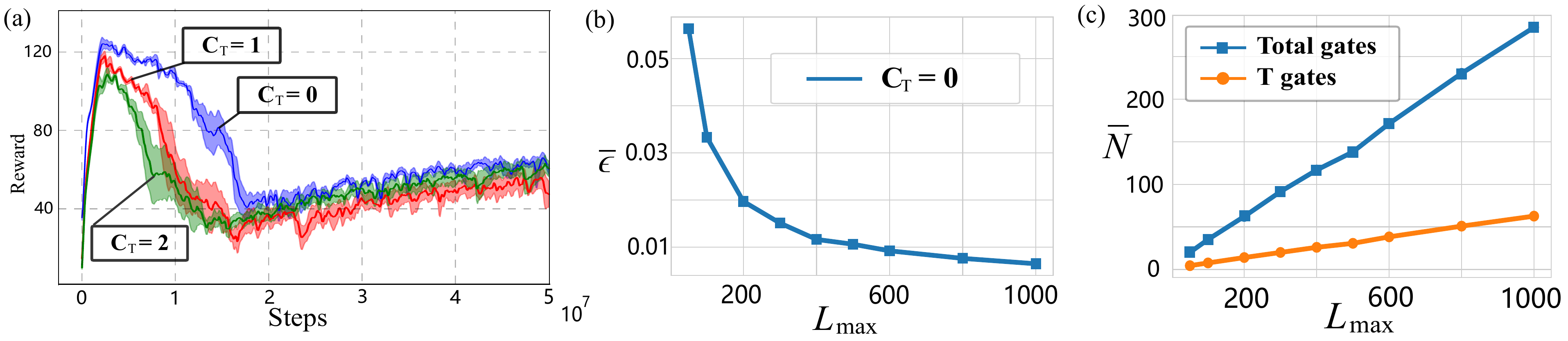}
    \caption{(a) The training process of the DRL agents for different $T$ cost $C_T=0,1,2$. We plot the average reward value as a function of the number of steps in the training process. (b) The relation of the average compiling error $\overline{\epsilon}$ over the test dataset with the maximal length $L_{\text{max}}$ of generated sequences from the agent. The cost of $T$ gate is set to $C_T=0$. (c) The average length of the gate sequences and the average number of $T$ gates produced by the agent when we increase the maximal length $L_{\text{max}}$ of generated sequences. The cost of $T$ gate is set to $C_T=0$.}
    \label{fig:SupDRL}
\end{figure*}

In this section, we provide the design details of the agent and algorithm we used. We additionally provide more numerical data about applying this algorithm to compilation based on Majorana fermion systems.  

\subsection*{I. Training the PPO agent}
Our DNN provided by \textsc{OPENAI} baseline package \cite{Brockman2016Openai} consists of five full connected layers each containing $256$ neurons. The activate function is the leaky ReLU function \cite{Maas2013Rectifier} throughout the neural network. We exploit the Adam algorithm \cite{Kingma2014Adam,Sashank2018Convergence} as our optimizer, and batch normalization is applied.

As mentioned in the main text, the DNN is trained to evaluate the objective function $J(\pi)$ with reward function:
\begin{equation}\label{eq:suppreward_all}\tag{D1}
r_n=r_s(U_n, U_t) - C_{g^*},
\end{equation}
\begin{equation}\label{eq:suppreward}\tag{D2}
r_s=\begin{cases}
c(1+\max\{0, 1-\frac{n}{L_{U_t}+10}\}),&d(U_n,U_t)<\epsilon_t,\\
\frac{-d(U_n,U_t)}{L_{\text{max}}},&\text{Otherwise},
\end{cases}
\end{equation}
where $r_s(U_n,U_t)$ is the state reward obtained by comparing the distance $d(U_n,U_t)$ between the approximation gate $U_n$ and the target gate $U_t$, $C_{g^*}$ is the additional punishment for the employment of $g^*$ gate, $L_{U_t}$ is the number of gates used for generating $U_t$, $L_{\text{max}}$ is the maximal number of steps allowed to compile the target gate for the agent, $c$ is a constant to balance rewards and punishments, and $\epsilon_t$ is the distance tolerance.

Starting from the identity at each iteration, the agent chooses a gate from $A=\{B_{12},B_{12}^{-1},B_{23},B_{23}^{-1},T, T^{-1}\}$ in each step and obtains a reward value calculated by Eq.~\eqref{eq:suppreward_all}. When the distance between the approximation sequence and the target unitary gate falls within the threshold $\epsilon_t$, the agent obtains a reward and starts a new iteration with a new target gate. When the number of the steps exceeds the maximal length $L_{\text{max}}$, the iteration also terminates.

Before the training process, the DNN is initialized with random parameters. From the beginning of the training process, we feed random sequences consisting gates from $A=\{B_{12},B_{12}^{-1},B_{23},B_{23}^{-1},T, T^{-1}\}$ of length $10$. We choose the accuracy threshold $\epsilon_t=10^{-3}$ and train the agent and the DNN to search for $\pi$ with higher reward and generate approximations below the maximal length $L_{\text{max}}$. During the training process, we hold a reward threshold as a function of the length of the random sequence in the training data. If the reward obtained by the agent when compiling the training data reaches a threshold, we increase the length of the random sequence generated as training data until the length reaches $80$. In Fig. \ref{fig:SupDRL}(a), we plot the average reward as a function of the number of steps of training. We can observe that the average reward first increases to the reward threshold and keeps dropping when we increase the length of the sequence among training data. When the length of the randomly generated training sequence reaches the upper bound $80$, the average reward would increase as we no longer increase the length. Moreover, we could obtain that the average reward would decrease if we increase the cost for $T$ gate. We trained this model on a single NVIDIA TITAN V GPU for about one day.

\subsection*{II. More results on applying PPO algorithms to topological quantum compiling on Majorana fermions}
To further explore our DRL algorithm in compiling topological quantum compiling on Majorana fermions, we construct another test dataset consisting of $1000$ random generated sequences of length $80$ with gates chosen from $A=\{B_{12},B_{12}^{-1},B_{23},B_{23}^{-1},T, T^{-1}\}$. We feed this dataset into the trained agent and increase the maximal length $L_{\text{max}}$ of generated sequences to observe the changes of average distance $\overline{\epsilon}$, number of $T$ gates, and approximation sequence length.

As shown in Fig. \ref{fig:SupDRL}(b), we input the test dataset into the agent trained with $C_T=0$. We observe that when the maximal length $L_{\text{max}}$ increases, the average compilation error first decreases quickly and then converges to a stable value of about $0.005$. This indicates that when $L_{\text{max}}$ exceeds a threshold, the major obstacle for improving the accuracy of the compiler would be the sparsity of the net with approximated policy reward of the agent. It is worthwhile to mention that compare with Ref. \cite{Zhang2020Topological}, the search complexity in our algorithm increases linearly rather than exponentially with $L_{\text{max}}$. In Fig. \ref{fig:SupDRL}(c), we plot the average length of the approximation sequences and the number of $T$ gates in the sequences as a function of $L_{\text{max}}$. It is shown that these two functions increase approximately linearly with $L_{\text{max}}$ while the proportion of $T$ gate in the approximation sequence remains rarely changed. This result indicates that our DRL agent could stably reduce the usage of $T$ gate under different $L_{\text{max}}$. 

\section*{E. Encoding and operation on Majorana fermions}
In this section, we briefly introduce the encoding methods on Majorana fermions systems. The fusion principle for Majorana fermions is the Ising type $\tau\times\tau\sim \mathbf{I}+\psi$ with $\mathbf{I},\tau,\psi$ representing a vacuum state, a Majorana fermion, and a normal fermion. In the main text, we consider the four-quasiparticle encoding scheme where each qubit is encoded by four Majorana fermions with the total topological charge as $0$. The logical basis states for the qubit are $|0\rangle_L=|[(\bullet,\bullet)_\mathbf{I},(\bullet,\bullet)_\mathbf{I}]_{\mathbf{I}}\rangle$ and $|1\rangle_L=|[(\bullet,\bullet)_\psi,(\bullet,\bullet)_\psi]_{\mathbf{I}}\rangle$. Here, each $\bullet$ is a Majorana fermion, and $\mathbf{I},\psi$ are the two possible fusion channels of a pair of Majorana fermions. 

As mentioned in the main text, the gates \{H,S\} could be realized by braidings on the four-quasiparticle scheme. We denote the four Majorana braiding operators on each quasiparticle as $b_i,i=1,2,3,4$ in one logic qubit and these operators satisfy $b_i^{\dagger}=b_i,b_i^2=I$ and anticommutation relation $\{b_i,b_j\}=2\delta_{ij}$. As shown in Ref. \cite{Hassler2010Anyonic}, Pauli operators in computational basis can be expressed as
\begin{equation}\label{eq:Paulimajorana}\tag{E1}
\sigma^x=-ib_2 b_3,\sigma^y=-ib_1b_3,\sigma^z=-ib_1b_2.
\end{equation}

Unitary operations can be realized by counterclockwise exchanges of two Majorana fermions as below:
\begin{equation}\label{eq:Unitarymajorana}\tag{E2}
B_{jj'}=e^{i\pi/4(ib_j b_{j'})},
\end{equation}
where $j,j'$ are chosen as two neighboring quasiparticles. Specifically, we give three basic braidings as
\begin{equation}\label{eq:braidmajorana}\tag{E3}
B_{12}=B_{34}\cong\begin{pmatrix}1 & 0\\0 & i\end{pmatrix}=S,B_{23}\cong\frac{1}{\sqrt{2}}\begin{pmatrix}1 & -i\\-i & 1\end{pmatrix}.
\end{equation}

We can implement H gate with the braiding sequence $H=B_{23}^2B_{12}^{-1}B_{23}B_{12}^{-1}B_{23}^2$. Hence, we have shown that a single-qubit Clifford group generated by H gate and S gate can be realized by Majorana braidings. However, entanglement gate on two-qubit cannot be obtained through braiding due to the no entanglement rule \cite{Bravyi2006Universal}.

In order to implement a two-qubit control gate, we introduce an accessory topological manipulation called nondestructive measurement of the anyon fusion \cite{Bravyi2005Universal,bravyi2002Fermionic}, which can be implemented through the anyon interferometry. We denote the eight Majorana modes on the two logical qubits as $b_1,...,b_8$, where the control (target) qubit are encoded by the first (last) four modes, respectively. The two-qubit controlled phase flip gate $\Pi(\sigma^z)$ can be represented by:
\begin{equation}\label{eq:CZmajorana}\tag{E4}
\Pi(\sigma^z)=e^{-(\pi/4)b_3b_4}e^{-(\pi/4)b_5b_6}e^{(i\pi/4)b_3b_4b_5b_6}e^{i\pi/4}.
\end{equation}

In the representation above, an ancillary pair $b_9b_{10}$ is added. We measure the fusion of the four Majorana modes $b_4b_3b_6b_9$ and get outcome $\pm1$, which corresponds to the vacuum state and the normal fermion with projectors $\Lambda_{\pm}^{(4)}=\frac12(1\pm b_4b_3b_6b_9)$. Then, we can measure fusion of the Majorana modes (operator) $-ib_5b_9$ with similar method and get projectors $\Lambda_{\pm}^{(2)}=\frac12(1\mp ib_5b_9)$. We have the following relation
\begin{equation}\label{eq:fusionmeasuremajorana}\tag{E5}
e^{(i\pi/4)b_3b_4b_5b_6}=2\sum_{\kappa,\xi=\pm}V_{\kappa\xi}\Lambda^{(2)}_{\kappa}\Lambda^{(4)}_{\xi},
\end{equation}
where $V_{++}=e^{(\pi/4)b_5b_{10}}$, $V_{--}=e^{-(\pi/4)b_5b_{10}}$, $V_{+-}=i e^{(\pi/2)b_4b_3}e^{(\pi/2)b_5b_6}e^{(\pi/4)b_5b_{10}}$, $V_{-+}=i e^{(\pi/2)b_4b_3}e^{(\pi/2)b_5b_6}e^{-(\pi/4)b_5b_{10}}$ can be implemented by one or several braiding operations of Majorana fermions.

\end{document}